\documentclass[preprint,proceedings]{rmaa}

\newcommand{\izw}{I\,Zw\,18}
\newcommand{\sbs}{SBS\,0335-052}

\newcommand{\D}{\discretionary{}{}{}}

\SetYear{2002}
\SetConfTitle{Galaxy Evolution: Theory and Observations}

\title{Low-Metallicity Blue Compact Dwarfs as Templates for Primordial Star Formation}

\author{
  L.~K.~Hunt\altaffilmark{1},
  H.~Hirashita\altaffilmark{2},
  T.~X.~Thuan\altaffilmark{3},
  Y.~I.~Izotov\altaffilmark{4},
  L.~Vanzi\altaffilmark{5} }

\altaffiltext{1}{Istituto di Radioastronomia-Sez. Firenze/CNR,
  Largo E. Fermi 5, 50125 Firenze, Italy
  (hunt\D{}@arcetri.\D{}astro.\D{}it).}
\altaffiltext{2}{INAF-Osservatorio Astrofisico di Arcetri, Firenze, Italy.}
\altaffiltext{3}{University of Virginia, Charlottesville, VA, USA.}
\altaffiltext{4}{Main Astronomical Observatory, Kyiv, Ukraine.}
\altaffiltext{5}{European Southern Observatory, La Silla, Chile.}

\suppressfulladdresses

\listofauthors{L. K.~Hunt, H.~Hirashita, T.~X.~Thuan, Y.~I.~Izotov, \& L.~Vanzi}
\indexauthor{Hunt, L. K.}
\indexauthor{Hirashita, H.}
\indexauthor{Thuan, T. X.}
\indexauthor{Izotov, Y. I.}
\indexauthor{Vanzi, L.}

\addkeyword{Galaxies: compact}
\addkeyword{Galaxies: evolution}
\addkeyword{Galaxies: ISM}
\addkeyword{Galaxies: starburst}
\addkeyword{Galaxies: stellar content}

\begin{document}
\maketitle 

\boldabstract{Understanding how galaxies formed their first stars is a 
vital cosmological question, but the study of high-redshift objects,
caught in the act of forming their first stars, is difficult.
Here we argue that two extremely low-metallicity Blue Compact
Dwarf galaxies (BCDs), \izw\ and \sbs, could be local templates for primordial 
star formation, since both lack evolved ($>\,$1\,Gyr) stellar populations;
but they form stars differently.
}

The main body of \izw\ (1/50\,$Z_\odot$)
comprises two star-forming complexes (SFCXs), in the NW
and the SE, which are surrounded by a smooth red envelope.
Our new deep near-infrared images combined
with HST/WFPC2 $BVI$ show that the red extended region is significantly
contaminated by ionized gas emission (Hunt et al. 2002).
Such emission reddens the observed colors, mimicking the effect
of age.
Indeed, an analysis of five colors ($B-H$, $V-K$, $V-I$, $J-H$, $H-K$) 
in terms of stellar populations, extinction, and gas,
suggests that the oldest stars in the outskirts
of \izw\ (and in the detached C component) can be no older than 500\,Myr.

Unlike \izw, where there are no super-star clusters (SSCs), 
virtually all of the star formation in \sbs\ (1/40\,$Z_\odot$)
occurs in them.
The underlying extended envelope is 
dominated by ionized gas emission rather than old stars (Vanzi et al. 2000).
There is also evidence (Thuan et al. 1999, Hunt et al. 2001) for
$A_V=15$\,mag extinction, caused by $10^4 - 10^5$ $M_\odot$
of dust.
Our Br$\alpha$ measurement suggests that more than 3/4 of the star formation
in \sbs\ is obscured by dust, not clearly visible even at 2\,$\mu$m. 

\begin{figure}[!t]
  \includegraphics[width=\columnwidth]{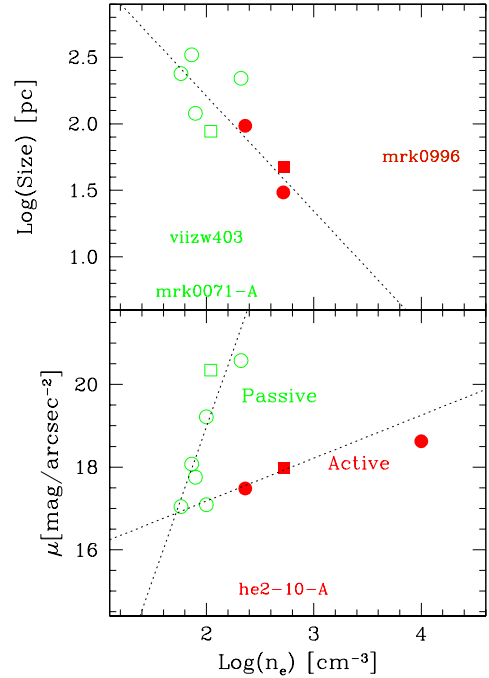}
  \vspace{-0.9cm}
  \caption{
Size (top panel) and surface brightness (bottom) of the
SFCXs plotted against the electron density: 
\izw\ (open square); \sbs\ (filled square).
  \label{fig:p4}}
\end{figure}

Taken together, these results suggest that 
\izw\ and \sbs\ form stars in very different ways,
in spite of their similar metallicity and lack of old stars.
We propose that dust and SSCs are associated with an ``active''
mode of star formation, and the lack of them with a ``passive'' mode.
Theoretical models suggest that the dust content of \sbs\ is
consistent with Type II SNe production, and that 
{\it high density} and {\it compact size} of
SFCXs favor enhancement of dust shielding in
chemically unevolved galaxies (Hirashita et al. 2002).
We have investigated this observationally by
measuring the sizes of SFCXs in a small sample
of BCDs observed with HST/WFPC2 (see Fig. \ref{fig:p4}).
It turns  out that size and density (shown in the top panel) are
significantly ($>$\,99\%) correlated, once we eliminate the uncertain density
estimates (VII\,Zw\,403, Mrk\,0071-A, Mrk\,996).
This correlation suggests that compact size and high density could
also be hallmarks of the {\it active} regime.

\begin{thebibliography}

\bibitem{hirashita} Hirashita,H., Hunt,L.K., Ferrara,A. 2002,
MNRAS, 330,L19

\bibitem{hunt02} Hunt,L.K., Thuan,T.X., Izotov,Y.I. 2002,ApJ,submitted

\bibitem{hunt01} Hunt,L.K., Vanzi,L., Thuan,T.X. 2001, A\&A, 377,66

\bibitem{thuan} Thuan,T.X., Sauvage,M., Madden,S. 1999, ApJ, 516,753

\bibitem{vanzi} Vanzi,L. et al. 2000, A\&A, 363,493

\end{thebibliography}
\end{document}